\documentclass[
  prl,
  twocolumn,
  twoside,
  byrevtex,
  superscriptaddress,
  nofootinbib,
  floatfix,
  longbibliography,
]{revtex4-2}

\usepackage[realmainfile]{currfile}
\input{\currfilebase.settings}
\usepackage{environ}
\usepackage{ragged2e}

\usepackage[export]{adjustbox}

\usepackage{colortbl}

\PassOptionsToPackage{hyphens}{url}\usepackage{hyperref}
\makeatletter
\g@addto@macro{\UrlBreaks}{\UrlOrds}
\makeatother

\hypersetup{
  colorlinks=true,
  allcolors=todoblue,
  urlcolor=todoblue,
  citecolor=todoblue,
  pdfborder={0 0 0},
  breaklinks=true,
}

\usepackage[normalem]{ulem}

\usepackage{soul}

\makeatletter
\let\ftype@table\ftype@figure
\makeatother

\usepackage{makecell}

\usepackage{array}

\usepackage{color}

\definecolor{todoblue}{RGB}{0, 91, 187}

\usepackage{tocloft}

\newenvironment{textblock}{\renewcommand{\item}{}\ignorespaces}{}

\usepackage{changepage}

\usepackage{graphicx}
\usepackage{verbatim}
\usepackage{amssymb}
\usepackage{amsmath}
\usepackage{ifthen}

\usepackage{longtable}

\usepackage{xcolor}

\definecolor{olivegreen}{rgb}{0.33333,.41961,0.18431}
\definecolor{forestgreen}{rgb}{0.13333,.5451,0.13333}

\definecolor{lightgrey}{rgb}{0.7,0.7,0.7}
\definecolor{verylightgrey}{rgb}{0.90,0.90,0.90}
\definecolor{veryverylightgrey}{rgb}{0.95,0.95,0.95}
\definecolor{grey}{rgb}{0.5,0.5,0.5}
\definecolor{darkgrey}{rgb}{0.3,0.3,0.3}
\definecolor{verydarkgrey}{rgb}{0.15,0.15,0.15}

\definecolor{headerblue}{HTML}{33367E}
\definecolor{unitednationsblue}{HTML}{4D88FF}

\definecolor{charcoal}{HTML}{36454F}
\definecolor{cinerous}{HTML}{98817B}
\definecolor{feldgrau}{HTML}{4D5D53}
\definecolor{glaucous}{HTML}{6082B6}
\definecolor{arsenic}{HTML}{3B444B}
\definecolor{xanadu}{HTML}{738678}

\definecolor{firebrick}{HTML}{B22222}
\definecolor{orangered}{HTML}{FF4500}
\definecolor{tomato}{HTML}{FF6347}

\definecolor{orange}{RGB}{255,116,0}

\definecolor{purpletaupe}{HTML}{3B444B}

\definecolor{rose}{HTML}{E3242B}

\colorlet{editnotecolor}{rose}

\definecolor{headerorange}{RGB}{255,116,0}
\definecolor{headergray}{RGB}{230,230,230}

\definecolor{headerpop}{RGB}{230,230,230}

\definecolor{magmalight}{RGB}{252,251,195}
\definecolor{magmalightalt}{RGB}{250,240,184}
\definecolor{magmamedium}{RGB}{245,200,146}
\definecolor{magmadark}{RGB}{224,106,98}

\definecolor{icelight}{RGB}{223,242,244}
\definecolor{icelightalt}{RGB}{189,222,226}
\definecolor{icemedium}{RGB}{132,184,204}
\definecolor{icedark}{RGB}{103,153,191}

\definecolor{traitrowcolor}{RGB}{223,242,244}
\definecolor{traitrowcoloralt}{RGB}{189,222,226}

\definecolor{characterrowcolor}{RGB}{252,251,195}
\definecolor{characterrowcoloralt}{RGB}{250,240,184}

\definecolor{archetyperowcolor}{RGB}{255,213,212} 
\definecolor{archetyperowcoloralt}{RGB}{255,182,179} 

\definecolor{datasetrowcolor}{RGB}{232,244,234}
\definecolor{datasetrowcoloralt}{RGB}{210,231,214}

\NewEnviron{excerpt}{
    \begin{quote}
      \medskip
      \BODY
      \medskip
    \end{quote}
}

\NewEnviron{editnote}{
    \begin{quote}
      \color{rose}
      $\blacksquare$
      \BODY
      \medskip
    \end{quote}
}

\newcommand{\command}[1]{
  \lstinline[language={[LaTeX]TeX},basicstyle=\ttfamily]{#1}
}

\newcommand{\editbox}[2]{
}

\newcommand{\editboxwithlatex}[2]{
}

\usepackage{fancyvrb}

\usepackage{tcolorbox}
\tcbuselibrary{skins,breakable,listings,breakable}

\usepackage{tikz}
\usetikzlibrary{shapes}

\tikzstyle{mybox} = [draw=lightblue!70, fill=lightblue!7, very thick,
    rectangle, rounded corners, inner sep=10pt, inner ysep=20pt]

\tikzstyle{editortitle} =[draw=archetyperowcoloralt, fill=archetyperowcoloralt, text=black]

\newcommand\Loadedframemethod{default}
\usepackage[framemethod=\Loadedframemethod]{mdframed}

\mdfsetup{skipabove=\topskip,skipbelow=\topskip}

\tikzstyle{loglinetitle} =[draw=icedark, fill=icemedium!50, text=black]

\tikzstyle{abstracttitle} =[draw=magmadark!75, fill=magmamedium!75, text=black]

\tikzstyle{infotitle} =[draw=darkgrey, fill=lightgrey!50, text=black]

\usepackage{enumitem}

\tikzstyle{changelogtitle} =[draw=darkgrey, fill=lightgrey!50, text=black]

\usepackage{listings}

\usepackage{stringstrings}
\usepackage{readarray}

\def\firstchar#1#2|{#1}
\edef\tbs{\detokenize{\X}}
\edef\tbs{\expandafter\firstchar\tbs|}
\edef\tlb{\detokenize{{}}}
\edef\tlb{\expandafter\firstchar\tlb|}
\edef\tus{\detokenize{_}}
\newcounter{index}
\newcommand\detokenizeplus[1]{%
  \def\temparg{\detokenize{#1}}%
  \getargsC{\temparg}%
  \setcounter{index}{0}%
  \def\prevmacro{F}%
  \whiledo{\value{index} < \narg}{%
    \stepcounter{index}%
    \isnextbyte[q]{\tbs}{\csname arg\roman{index}\endcsname}%
    \if T\theresult%
      \if T\prevmacro\unskip\else\fi%
      \def\prevmacro{T}%
    \else%
      \def\prevmacro{F}%
   \fi%
    \isnextbyte[q]{\tlb}{\csname arg\roman{index}\endcsname}%
    \if T\theresult\unskip\else\fi%
    \isnextbyte[q]{\tus}{\csname arg\roman{index}\endcsname}%
    \if T\theresult\unskip\else\fi%
    \csname arg\roman{index}\endcsname~%
  }%
}

\usepackage{mathtools}

\newboolean{twocolswitch}

\newcommand{\sindex}[1]{}
\newcommand{\nindex}[1]{}

\newcommand{\etal}{\textit{et al.}}
\newcommand{\www}[1]{\url{#1}}

\usepackage{lettrine}

\usepackage{etoolbox,refcount}

\newcounter{countitems}
\newcounter{nextenumeratecount}
\newcommand{\setupcountitems}{%
  \stepcounter{nextenumeratecount}%
  \setcounter{countitems}{0}%
  \preto\item{\stepcounter{countitems}}%
}
\makeatletter
\newcommand{\computecountitems}{%
  \edef\@currentlabel{\number\c@countitems}%
  \label{countitems@\number\numexpr\value{nextenumeratecount}-1\relax}%
}
\newcommand{\nextenumeratecount}{%
  \getrefnumber{countitems@\number\c@nextenumeratecount}%
}
\makeatother
\AtBeginEnvironment{enumerate}{\setupcountitems}
\AtEndEnvironment{enumerate}{\computecountitems}

\usepackage{xfrac}

\newcommand{\dee}[1]{\textnormal{d}#1}

\newcommand{\rank}{r}
\newcommand{\rgrtime}{t}

\newcommand{\rankdistexponent}{\alpha}

\newcommand{\typesize}{S}

\newcommand{\growthfactorsymbol}{G}
\newcommand{\growthfactor}{\growthfactorsymbol_{\rgrtime,\rankdistexponent}}

\newcommand{\Ndistincttypes}{\numbersymbol_{\rgrtime,\rankdistexponent}}
\newcommand{\Ndistincttypesfn}[2]{\numbersymbol_{#1,#2}}

\newcommand{\LambertW}{W}

\newcommand{\genharmonicsum}[2]{H_{#1}^{(#2)}}

\newcommand{\numbersymbol}{N}

\setboolean{twocolswitch}{true}

\begin{document}




\raggedright

\title{\protect
  Complete asymptotic type-token relationship for
\\
growing complex systems
with inverse power-law count rankings






}

\author{
\firstname{Pablo}
\surname{Rosillo-Rodes}
}
\email{prosillo@ifisc.uib-csic.es}

\affiliation{
     Institute for Cross-Disciplinary Physics and Complex Systems IFISC (UIB-CSIC), Campus Universitat de les Illes Balears, E-07122 Palma de Mallorca, Spain
}

\affiliation{
  Computational Story Lab,
  University of Vermont,
  Burlington,
  VT 05405,
  US
}

\author{
\firstname{Laurent}
\surname{H\'ebert-Dufresne}
}

\email{laurent.hebert-dufresne@uvm.edu}

\affiliation{
  Department of Computer Science,
  Vermont Complex Systems Institute,
  MassMutual Center of Excellence for Complex Systems and Data Science,
  Vermont Advanced Computing Center,
  University of Vermont,
  Burlington,
  VT 05405,
  US
}

\affiliation{
  Santa Fe Institute,
  1399 Hyde Park Rd,
  Santa Fe,
  NM 87501,
  US
}

\author{
\firstname{Peter Sheridan}
\surname{Dodds}
}
\email{peter.dodds@uvm.edu}

\affiliation{
  Computational Story Lab,
  University of Vermont,
  Burlington,
  VT 05405,
  US
}

\affiliation{
  Department of Computer Science,
  Vermont Complex Systems Institute,
  MassMutual Center of Excellence for Complex Systems and Data Science,
  Vermont Advanced Computing Center,
  University of Vermont,
  Burlington,
  VT 05405,
  US
}

\affiliation{
  Santa Fe Institute,
  1399 Hyde Park Rd,
  Santa Fe,
  NM 87501,
  US
}

\date{\today}

\begin{abstract}
  \protect
  \begin{textblock}
\item 
The growth dynamics of complex systems often exhibit statistical regularities involving power-law relationships. 
\item
For real finite complex systems formed by countable tokens (animals, words)
as instances of distinct types (species, dictionary entries),
an inverse power-law scaling 
$\typesize 
\sim 
\rank^{-\rankdistexponent}$
between type count $\typesize$
and type rank $\rank$, widely known as Zipf's law, is widely observed to varying degrees of fidelity.
\item
A secondary, summary relationship is Heaps' law,
which states that the number of types
scales sublinearly with the total number of observed tokens
present in a growing system.
\item
Here, we propose an idealized model
of a growing system that
(1) 
deterministically produces arbitrary inverse power-law 
count rankings for types,
and
(2) 
allows us to determine the exact asymptotics of
the type-token relationship.
\item 
Our argument improves upon and remedies earlier work.
\item
We obtain a unified asymptotic expression for all
values of $\rankdistexponent$,
which corrects the special cases of
$\rankdistexponent = 1$
and
$\rankdistexponent \gg 1$.
\item
Our approach relies solely on the form of count rankings,
avoids unnecessary approximations,
and does not involve any stochastic mechanisms or sampling processes.
\item
We thereby demonstrate that a general type-token relationship arises solely as a consequence of Zipf's law.
\end{textblock}


 
\end{abstract}

\maketitle




\justifying 



\begin{textblock}
\item 
  \textit{Introduction}---Universal statistical regularities play a central role in the study of complex systems.
\item 
  Prominent examples are power-law relationships, 
  such as Zipf’s law for word frequencies~\cite{zipf1949a,newman2005b},
\item 
  observed across human~\cite{altmann2016} and animal~\cite{arnon2025} communication. 
\item 
  For language, Zipf’s law states that the count of a word
  in a text decreases with its rank $r$ as
  $S_{\rank,\rankdistexponent} \sim r^{-\rankdistexponent}$, 
  originally with $\rankdistexponent \simeq 1$.
\item 
  While arguments about 
  the empirical validity and mechanistic origins 
  of Zipf's law have remained contentious for the 
  better part of a century~\cite{yule1925,mandelbrot1953a,simon1955a,maillart2008a,ferrericancho2010a,williams2015b},
\item 
  count-rankings adhering approximately to an inverse power-law have been observed across systems 
  of all kinds including
  ecological systems~\cite{yule1925, camacho2001}, 
  complex networks~\cite{Redner1998, barabasi1999},  
  and socioeconomic systems~\cite{serrano2003,pinto2012, gabaix2016,gaillard2023}. 
\end{textblock}

\begin{textblock}
\item 
  Another well-known scaling regularity is Heaps’
  law~\cite{herdan1960,heaps1978,Chacoma2020},  
\item
  which describes how the vocabulary count $N$ of a text grows with its total length $t$ as $N_{t, \beta} \sim t^{\beta}$,  with $0 < \beta < 1$. 
\item
  Like Zipf's law,
  type-token scaling has been found to generalize beyond language,
  and has been reported to hold in, for example,
\item
  chemoinformatics~\cite{benz2008},
  computer code~\cite{zhang2009},
  and urban systems~\cite{simini2019}.
\item 
 The type-token relationship is a summary one,
 as it carries far less information than count ranking
 which records not just type ranks but the names
 of the types themselves. 
\end{textblock}

\begin{textblock}
\item 
The fact that Zipf’s and Heaps’ laws are empirically observed to hold simultaneously has motivated interest in whether or not a fundamental relationship between them exists in finite, real systems. 
\item
Previous works have developed a range of frameworks to relate both scaling laws, for example, by
\item
empirical observation~\cite{baeza-yates2000}, 
\item
proposing specific language models~\cite{zanette2005,serrano2009}, 
\item
growth dynamics~\cite{laurent2016},
\item
sampling mechanisms~\cite{vanleijenhorst2005, eliazar2011}, 
\item
or assuming in advance their simultaneous validity~\cite{bernhardsson2009}.
\item
L\"{u} \textit{et al.}~\cite{lu2010a}, whose work we improve upon,
found
that the scaling relation between the two exponents in the infinite system count
limit is given by
\item
\begin{equation}
    \beta = 
    \begin{cases}
        1
        & 
        \textnormal{for} 
        \ 
        \alpha \le 1,
        \\
             1/\alpha
        & 
        \textnormal{for} 
        \ 
        \alpha > 1.
    \end{cases}
    \label{eq:simple-type-token.lu-exponents}
\end{equation}
\end{textblock}

\begin{textblock}
\item
In this work, we derive a complete asymptotic type–token relationship for idealized growing finite systems, valid for any value of the scaling exponent 
$0 \leq \alpha < \infty$.
\item
We employ the Euler-Maclaurin expansion based on a simple growth model to improve the description of the $\alpha = 1$ and $\alpha \gg 1$ regimes. Our formulation corrects earlier approximations
and returns the scaling form of Heaps' law in specific limits.
\item
Our approach is independent of the underlying mechanisms of the system, provides an excellent fit to our growing system model, and offers further insight into the growth dynamics of real systems.
\end{textblock}


\begin{textblock}
\item
  \textit{Model of an idealized growing system}---We consider an idealized growing system comprising
  countable tokens as instances of types. 
\item
  We set aside physical mechanisms
  and consider a system emerging
  in time as follows.
\item
  The system grows in discrete time  
  so that there are $\rgrtime$ total
  tokens at time $\rgrtime = 1, 2, 3,   \ldots$,
  and such that the count of
  the $\rank$th type is
\item
  \begin{equation}
    \typesize_{\rank,\rgrtime,\rankdistexponent}
    =
    \left\lfloor
    \frac{1}{2}
    +
    \growthfactor
    \rank^{-\rankdistexponent}
    \right\rfloor,
    \label{eq:simple-type-token.Srt-base-equation-set-up}
  \end{equation}
\item
  where 
  $\rankdistexponent > 0$,
  $\lfloor \cdot \rfloor$
  is the floor operator, converted to a round-to-the-nearest-integer
  operator by the addition of 1/2, 
  and $\growthfactor$ is a growth factor
  which we define below.
\item
  We indicate count by $\typesize$ which stands
  for the more general conception of size.
\item
  For $\rankdistexponent = 0$,
  $
  \typesize_{\rank,\rgrtime,0}
  =
  1
  $
  for 
  $1 \le \rank \le t$
  and 
  0
  otherwise.
\item
  By construction, types remain ordered in count ranking
  according to their arrival time.
\item
  When the $\rank$th type's
  count
  $\typesize_{\rank,\rgrtime,\rankdistexponent}$
  first
  rounds up to 1 (i.e., reaches $1/2$),
  we may consider the type 
  as being created or uncovered.
\item
  At any time $t$, there will be a finite number
  $\Ndistincttypesfn{\rgrtime}{\rankdistexponent}$
  of
  types with one or more tokens in the system, $\Ndistincttypesfn{\rgrtime}{\rankdistexponent}$
  being also the rank of
  the most recently arrived type.
\item
  Because the system is a realized entity,
  we call 
  $\typesize_{\rank,\rgrtime,\rankdistexponent}$
  a count-ranking rather than a count-rank distribution,
  which mis-implies, for our model, a sampling process.
\end{textblock}

\begin{textblock}
\item 
  The growth factor
  $\growthfactor \ge 0$
  ``grows'' the system in a step-wise fashion so that
\item 
  \begin{equation}
    \rgrtime
    =
    \sum_{\rank=1}^{\infty}
    \left\lfloor
     \frac{1}{2}
     +
    \growthfactor
    \rank^{-\rankdistexponent}
    \right\rfloor.
    \label{eq:simple-type-token.Srt-base-equation-set-up-prefactor}
  \end{equation}
  
\item
 Equation~\eqref{eq:simple-type-token.Srt-base-equation-set-up-prefactor} means that, for a given value of $t$, there will be an increasingly small range of 
 values of $\growthfactor$ for which $t$ tokens are present in the system.
\item 
 Also, not all individual token counts $t$ are achievable, 
 because two types can reach a half integer
 simultaneously.
\end{textblock}

\begin{textblock}
\item
  Now, while our idealized system can be easily grown computationally,
  the floor operator prevents ready analysis.
\item
  We approximate the count-ranking of our model as
\item
  \begin{equation}
    \typesize_{\rank,\rgrtime,\rankdistexponent}
    =
    \left\{
    \begin{array}{ll}
      \typesize_{1,\rgrtime,\rankdistexponent}
      \rank^{-\rankdistexponent}
      &
      \textnormal{for}
      \quad
      1
      \le
      \rank
      \le
      \Ndistincttypes,
      \\
      0
      &
      \textnormal{for}
      \quad
      \rank
      >
      \Ndistincttypes.
    \end{array}
    \right.
    \label{eq:simple-type-token.Srt-approx}
  \end{equation}
  
\item
  Counts are now fractional,
  and we determine 
  $\typesize_{1,\rgrtime,\rankdistexponent}$ 
  below.
\item 
  We note that in some limits, the approximation may not allow
  us to capture the idealized system,
  particularly as $\alpha \rightarrow 0$, because the continuous form smooths over the discrete, step-wise growth that dominates when the count hierarchy flattens, causing the approximate scaling to fail to represent the actual ranked increments.
\item
  Nevertheless, we will be able to satisfactorily
  connect count-rank scaling
  to type-token scaling.
\end{textblock}

\begin{textblock}
\item
  To maintain the connection to our idealized base model, we enforce two conditions.
\item
  First, when a type first appears it has a count of 1.
\item  
  Setting rank $\rank$ to $\Ndistincttypes$
  in Eq.~\eqref{eq:simple-type-token.Srt-approx},
  we have
\item
  \begin{equation}
    \typesize_{\Ndistincttypes,t,\rankdistexponent}
    =
    1
    =
    \typesize_{1,\rgrtime,\rankdistexponent}
    \Ndistincttypes^{-\rankdistexponent},
    \label{eq:simple-type-token.S1-scaling-set-up}
  \end{equation}
\item
  which means
\item 
  \begin{equation}
    \typesize_{1,\rgrtime,\rankdistexponent}
    \sim
    \Ndistincttypes^{\rankdistexponent}.
    \label{eq:simple-type-token.S1-scaling}
  \end{equation}
\item
  Second, we must have the sum of fractional counts equal to time $\rgrtime$, so
\item 
  \begin{equation}
    \rgrtime
    =
    \sum_{\rank=1}^{N}
    \typesize_{\rank,\rgrtime,\rankdistexponent}
    \sim 
    \Ndistincttypes^\rankdistexponent 
    \genharmonicsum{\Ndistincttypes}{\rankdistexponent}
    \label{eq:simple-type-token.type-token-relationship}
  \end{equation}
\item 
where
\begin{equation}
    \genharmonicsum{n}{a} 
    =
    \sum_{k=1}^{n} 
    k^{-a}
\end{equation}
\item
is the $n$th generalized harmonic number
of order $a$.
\item
  Eq.~\eqref{eq:simple-type-token.type-token-relationship} implicitly gives us what we call
  the type-token relationship, i.e.,
  how the number of distinct types $\Ndistincttypes$ grows with the 
  overall number tokens, $\rgrtime$. 
\item 
    We may approximate the harmonic sum in Eq.~\eqref{eq:simple-type-token.type-token-relationship} by its integral form,
    \begin{equation}
    \genharmonicsum{\Ndistincttypes}{\rankdistexponent} 
    \sim 
    \int_{z=1}^{\Ndistincttypes}
    z^{-\rankdistexponent}
    \dee{z},
    \label{eq:simple-type-token.genharmonicsum-approx}
    \end{equation}
    so we would arrive at   
    \begin{equation}
        t \sim \frac{
      \Ndistincttypes^{\rankdistexponent}
    }{
      1 - \rankdistexponent
    }
    \left[
      \Ndistincttypes^{1-\rankdistexponent}
      -
      1
      \right].
      \label{eq:simple-type-token.integral_approximation_Lu}
    \end{equation}
    
\item    Equation~\eqref{eq:simple-type-token.integral_approximation_Lu} coincides with the result obtained by L\"{u}~\textit{et}~\textit{al.}\ in Ref.~\cite{lu2010a}, where the authors derive the connection between the number of distinct types and the total number of tokens by transforming their count-ranking into a size-frequency distribution (which has exponent $\gamma = 1 + 1/\rankdistexponent$) and approximating $\genharmonicsum{\Ndistincttypes}{\rankdistexponent}$ by Eq.~\eqref{eq:simple-type-token.genharmonicsum-approx}. 
\item
  Here, we have achieved a direct and compact derivation which proceeds exclusively from the count-rank representation.  
\end{textblock}

\begin{textblock}
\item 
  However, Eq.~\eqref{eq:simple-type-token.integral_approximation_Lu}
  is incorrect for large $\rankdistexponent$,
  and the error is in the 
  integral approximation of
  Eq.~\eqref{eq:simple-type-token.genharmonicsum-approx} (which we explain below).
\item 
    We return to Eq.~\eqref{eq:simple-type-token.type-token-relationship},
    and instead use the Euler-Maclaurin expansion~\cite{apostol1999} of $\genharmonicsum{\Ndistincttypes}{\rankdistexponent}$ for $\rankdistexponent \neq 1$, i.e.,
    \item 
    \begin{align}
        \genharmonicsum{\Ndistincttypes}{\rankdistexponent} 
        &
        =
        \frac{
        \Ndistincttypes^{1-\rankdistexponent}
        }{
        1-\rankdistexponent
        }
        +
        \zeta(\rankdistexponent)
        +
        \frac{1}{2} \Ndistincttypes^{-\rankdistexponent}
        \\
        \notag
        &-
        \frac{
        \rankdistexponent
        }{
        12 
        }
        \Ndistincttypes^{-\rankdistexponent-1}
        +
        \mathcal{O}\left(
        \Ndistincttypes^{-\rankdistexponent-2}\right).
    \end{align}
    
    \item
We now have
\item
    \begin{equation}
        \rgrtime \simeq 
        \frac{
        \Ndistincttypes
        }{
        1-\rankdistexponent
        }
        +
        \Ndistincttypes^\rankdistexponent \zeta(\rankdistexponent)
        +
        \frac{1}{2}
        -
        \frac{
        \rankdistexponent
        }{
        12 
        \Ndistincttypes
        } 
        +
        \mathcal{O}
        \left(
        \frac{
        1}{
        \Ndistincttypes^{2}
        }
        \right),
        \label{eq:simple-type-token.type-token-expansion-general-big}
    \end{equation}
    \item
which, to leading order, yields
\item
    \begin{equation}
        t
        \sim
        \frac{
        1
        }{
        1-\rankdistexponent
        }
        \Ndistincttypes
        +
        \zeta(\rankdistexponent)
        \Ndistincttypes^{\rankdistexponent},
        \label{eq:simple-type-token.type-token-expansion-general}
    \end{equation}
    \item
    where
    $\zeta(\rankdistexponent)$ is the Riemann zeta function for $\rankdistexponent > 1$ and its analytical continuation for $\rankdistexponent < 1$. 
\end{textblock}

\begin{figure*}[t]
\centering
  \includegraphics[width=\textwidth]{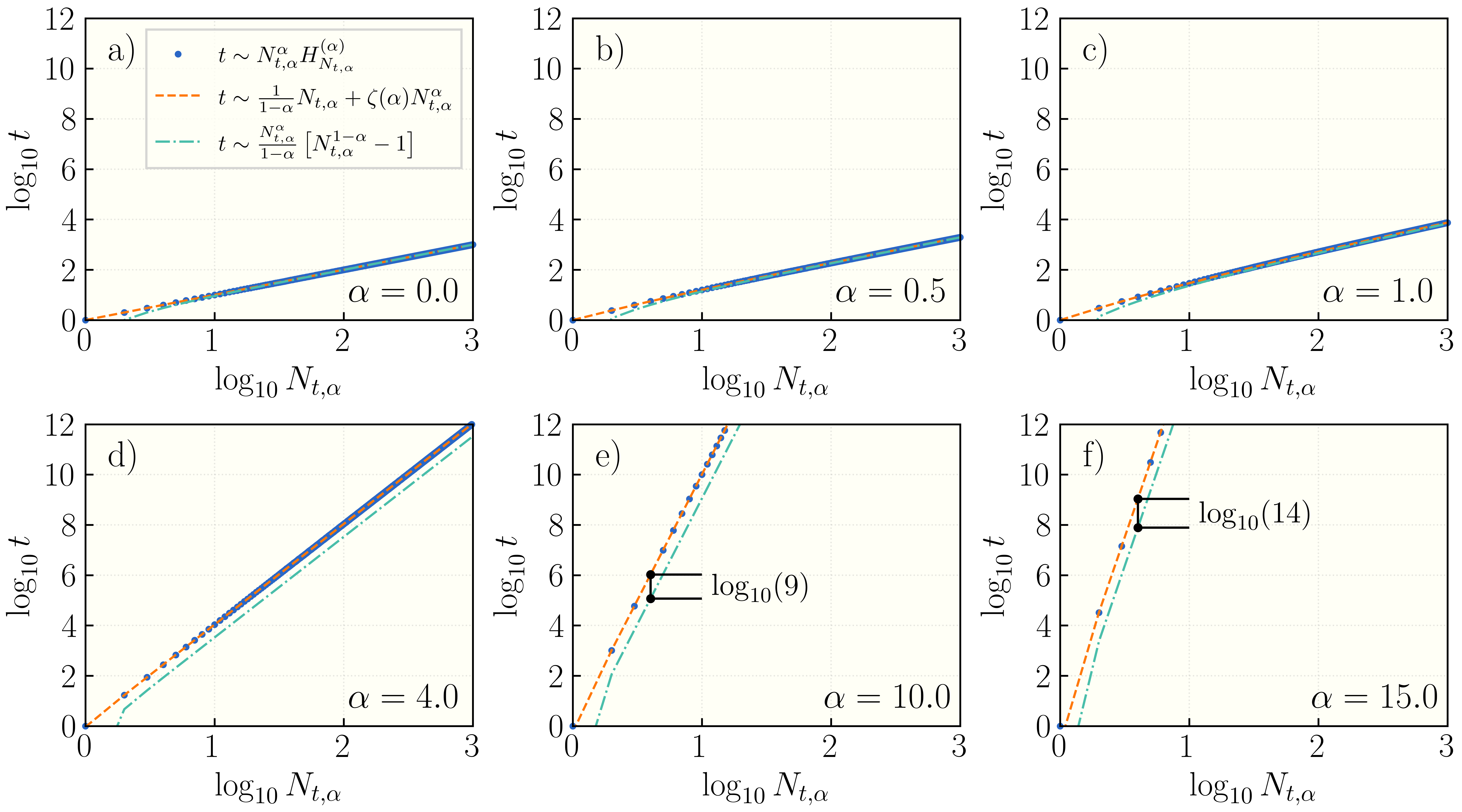}
    \caption{Comparison between data discretely generated using the harmonic sum in Eq.~\eqref{eq:simple-type-token.type-token-relationship} and the results yielded both by the expansion in Eq.~\eqref{eq:simple-type-token.type-token-expansion-general} and by Eq.~\eqref{eq:simple-type-token.integral_approximation_Lu} for different values of $\rankdistexponent$. In c), Eq.~\eqref{eq:simple-type-token.token-type-scaling-behavior} for $\alpha = 1$ is used to solve the divergence. In e) and f), we show the deviation of Eq.~\eqref{eq:simple-type-token.integral_approximation_Lu} from the data in the $\rankdistexponent \gg 1$ regime. 
    }
  \label{fig:simple-type-token.type_token_fit}
\end{figure*}

\begin{textblock}
    In panels Fig.~\ref{fig:simple-type-token.type_token_fit} a)--f), we show how Eq.~\eqref{eq:simple-type-token.type-token-expansion-general} fits the behavior of Eq.~\eqref{eq:simple-type-token.type-token-relationship} for five example values of $\rankdistexponent$. 
    \item
    As we show in the following, we can invert the limiting form of Eq.~\eqref{eq:simple-type-token.type-token-expansion-general} for certain values and ranges of $\rankdistexponent$ and compute $\Ndistincttypes(\rgrtime)$
    as a function of
    $\rgrtime$.
\end{textblock}

\begin{textblock}
\item
  \textit{Limiting behavior}---For large 
  $\rgrtime$ 
  and 
  $\Ndistincttypes$,
  we can express Eq.~\eqref{eq:simple-type-token.type-token-expansion-general} in simpler forms for
  the three regimes of 
  $\rankdistexponent \ll 1$,
  $\rankdistexponent \simeq 1$,
  and
  $\rankdistexponent \gg 1$.
  
\item
  First,
  for $\rankdistexponent \ll 1$, Eq.~\eqref{eq:simple-type-token.type-token-expansion-general-big} yields 
\item 
    \begin{equation}
    \rgrtime
    \sim
    \frac{1}{1-\rankdistexponent} \Ndistincttypes,
    \label{eq:simple-type-token.t-N-lpha<<1}
    \end{equation}
\item 
because
$\Ndistincttypes^{1} 
\gg 
\Ndistincttypes^{\alpha}
$
and
$
\lim\limits_{\rankdistexponent \rightarrow 0} \zeta(\rankdistexponent) 
 = 
 -1/2.
 $
\end{textblock}
 
\begin{textblock}
\item
Second, for $\rankdistexponent \rightarrow 1$, 
we use the approximations
\item
    \begin{equation}
        \Ndistincttypes^{\rankdistexponent} 
        \rightarrow 
        \Ndistincttypes 
        \left[
        1 
        - 
        (1 - \rankdistexponent) 
        \ln \Ndistincttypes 
        +
        \mathcal{O}
        \left(
        (1 - \rankdistexponent)^{2}
        \right)
        \right]
        \label{eq:simple-type-token.Nalpha-leading-order}
    \end{equation}
    \item
    and
    \item
    \begin{equation}
    (1 - \rankdistexponent)
    \zeta(\rankdistexponent) 
    \rightarrow 
    -1
     +
     \gamma_0
     (1 - \rankdistexponent)
     +
     \mathcal{O}
     \left(
     (
     1 - \rankdistexponent
     )^{2}
     \right)
     \label{eq:simple-type-token.zeta-leading-order}
    \end{equation}
    \item
    where
    $\gamma_{0} \simeq 0.577$ 
    is the Euler-Mascheroni constant. 
    \item
    Substituting
    Eqs.~\eqref{eq:simple-type-token.Nalpha-leading-order}
    and~\eqref{eq:simple-type-token.zeta-leading-order}
    into 
    Eq.~\eqref{eq:simple-type-token.type-token-expansion-general}, 
    we have
    for
    $\rankdistexponent
    \rightarrow 
    1$
    that
    \item
    \begin{equation}
        \rgrtime 
        \sim 
        \Ndistincttypes 
        \left(
        \ln 
        \Ndistincttypes 
        +
        \gamma_{0} 
        \right).
        \label{eq:simple-type-token.t-N-alpha=1}
    \end{equation} 
\item
    By next using the leading term of the asymptotic form of the Lambert $\LambertW$ function, we obtain
    \begin{equation}
        \Ndistincttypes 
        \sim 
        \frac{
        t
        }{
        \ln 
        \rgrtime
        + 
        \gamma_{0}
        - 
        \ln
    \left(
     \ln 
        \rgrtime
        + 
        \gamma_{0}
    \right)
        }.
    \end{equation}
\item 
Finally, for $\rankdistexponent \gg 1$, the exponential term in Eq.~\eqref{eq:simple-type-token.type-token-expansion-general}
dominates and because 
$\zeta(\rankdistexponent) \rightarrow 1$,
the limiting behavior is given by
\begin{equation}
\rgrtime
\sim
\Ndistincttypes^{\rankdistexponent}.
\end{equation}

\item
We observe that in the regime in which $\rankdistexponent \gg 1$, Eq.~\eqref{eq:simple-type-token.integral_approximation_Lu} 
incorrectly gives $t \rightarrow (\rankdistexponent-1)^{-1} \Ndistincttypes^{\rankdistexponent}$, 
which is off from the actual behavior of Eq.~\eqref{eq:simple-type-token.type-token-relationship} 
by a factor $(\rankdistexponent-1)^{-1}$.
\item
The reason for the error
is that the first term of 
$\genharmonicsum{\Ndistincttypes}{\rankdistexponent}$
is 1,
independent of $\rankdistexponent$,
but the steep decay for high $\rankdistexponent$
means that the integral approximation
of Eq.~\eqref{eq:simple-type-token.integral_approximation_Lu}
estimates it as 
$1/(\rankdistexponent-1)$.
\item
For steep enough power laws, as in panels e) and f) of Fig.~\ref{fig:simple-type-token.type_token_fit}
for $\rankdistexponent = 10$ and $\rankdistexponent = 15$, this represents nearly an order of magnitude underestimation of the system size $\rgrtime$ required to observe $\Ndistincttypes$ types, demonstrating that the integral approximation is insufficient for concentrated systems. There is evidence of physical systems that feature size-rankings with $\alpha >> 1$, such as music~\cite{perotti2020}, book sales, or personal wealth~\cite{newman2005}.
\item 
Note that we maintain $(\rankdistexponent-1)^{-1}$ instead 
of $\rankdistexponent^{-1}$ to match our 
simulations for these values of 
$\rankdistexponent$.
\end{textblock}

\begin{textblock}
\item
The above derived limiting forms of Eq.~\eqref{eq:simple-type-token.type-token-expansion-general} 
for 
large
  $\rgrtime$ 
  and 
  $\Ndistincttypes$,
can be summarized as
  \begin{equation}
    \rgrtime
    \sim
    \left\{
    \begin{array}{cl}
      \frac{1}{1-\rankdistexponent} \Ndistincttypes
      &
      \textnormal{for}
      \
      \rankdistexponent \ll 1,
      \\ 
      \Ndistincttypes 
      \left(
      \ln
      \Ndistincttypes 
      + 
      \gamma_{0} 
      \right)
      &
      \textnormal{for}
      \
      \rankdistexponent = 1,
      \\
      \Ndistincttypes^{\rankdistexponent}
      &
      \textnormal{for}
      \
      \rankdistexponent \gg 1,
      \\
    \end{array}
    \right.
    \label{eq:simple-type-token.token-type-scaling-behavior}
  \end{equation}
\item 
    and
  
\item
  \begin{equation}
    \Ndistincttypes
    \rightarrow
    \left\{
    \begin{array}{cl}
      (1-\rankdistexponent) \rgrtime
      &
      \textnormal{for}
      \
      \rankdistexponent \ll 1,
      \\
      \rgrtime/\left[
      \ln 
      \rgrtime
      + 
      \gamma_{0}
      -
      \ln
      \left(
     \ln 
        \rgrtime
        + 
        \gamma_{0}
    \right)
      \right]
      &
      \textnormal{for}
      \
      \rankdistexponent = 1,
      \\
      \rgrtime^{1/\rankdistexponent}
      &
      \textnormal{for}
      \
      \rankdistexponent \gg 1.
    \end{array}
    \right.
    \label{eq:simple-type-token.type-token-scaling-behavior}    
  \end{equation}
\end{textblock}


  

  
\begin{figure}[t]
\centering
  \includegraphics[width=\linewidth]{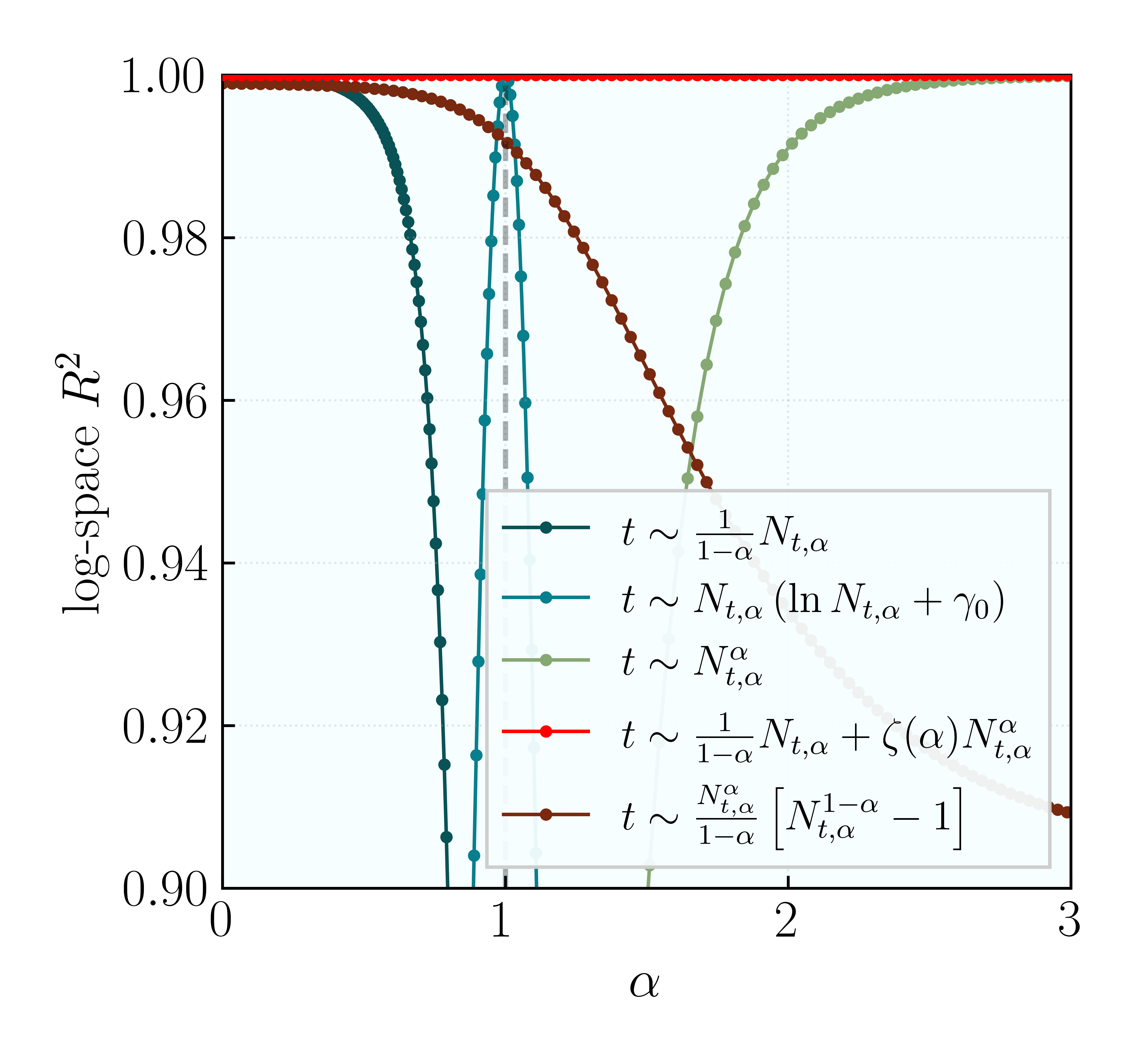}
    \caption{Goodness of the fit of the expansion in Eq.~\eqref{eq:simple-type-token.type-token-expansion-general} and the approximations in Eq.~\eqref{eq:simple-type-token.token-type-scaling-behavior}. The data used for the fit corresponds to $10^3$ points for $\Ndistincttypes \in [1, 10^3]$, computed using the sum in Eq.~\eqref{eq:simple-type-token.type-token-relationship}.
    }
  \label{fig:simple-type-token.type_token_fit_r2}
\end{figure}

\begin{textblock}
\item
  While Eq.~\eqref{eq:simple-type-token.type-token-expansion-general} fits the behavior correctly 
  for $\rankdistexponent\neq 1$, 
  the applicability of the simpler expressions for the limits in Eqs.~\eqref{eq:simple-type-token.token-type-scaling-behavior}~and~\eqref{eq:simple-type-token.type-token-scaling-behavior} depends on the range of $\rankdistexponent$, 
  as we demonstrate in Fig.~\ref{fig:simple-type-token.type_token_fit_r2}.
\end{textblock}

\begin{textblock}
\item 
\textit{Discussion}---In this work, we have presented a simple derivation of the asymptotic behavior of the type–token relationship for growing systems with inverse power-law count rankings.
\item
Our expression for the number of tokens $\rgrtime$ (equivalently system size)
as a function of the number of distinct types
$\Ndistincttypes$
in
Eq.~\eqref{eq:simple-type-token.type-token-expansion-general} 
fully captures large system behavior for all 
$\rankdistexponent \ge 0$.
\item 
Our results extend and correct those of 
\item
L\"{u} \etal~\cite{lu2010a}, 
\item 
achieving a more accurate approximation 
while only requiring count-ranking.
\item
Importantly, because our results follow from an idealized growing model,
Eq.~\eqref{eq:simple-type-token.type-token-expansion-general} is independent of the underlying mechanism that drives real system growth. 
\end{textblock}

\begin{textblock}
\item 
As can be visually appreciated in Fig.~\ref{fig:simple-type-token.type_token_fit}~a)--f), a linear approximation of the type–token relation in log–log space provides an accurate graphical description for any value of $\rankdistexponent$, provided that the system size $t$ is sufficiently large. 
\item
This observation, as discussed in Ref.~\cite{lu2010a}, 
clarifies why Heaps’ law so frequently appears in empirical studies.
\end{textblock}

\begin{textblock}
\item 
Our results demonstrate that Heaps' law is not an independent phenomenon, or one dependent on stochastic growth, but is rather an emergent statistical relationship of growing systems with 
inverse power-law count-rankings.
\end{textblock}


\bigskip

\acknowledgments

\textit{Acknowledgments}---
\unskip P.R. acknowledges support by the Spanish State Research Agency (MICIU/AEI/10.13039/501100011033) and FEDER (UE) under the Mar{\'\i}a de Maeztu project CEX2021-001164-M and
the projects PID2021-122256NB-C21 and PID2024-157493NB-C21, during a research stay in the Computational Story Lab at the University of Vermont.
The authors are grateful for
National Science Foundation Award \#2242829
(Science of Online Corpora, Knowledge, and Stories) and award \#2419733,
foundational support from MassMutual,
and
an anonymous philanthropic gift.



\bibliography{\filenamebase.bib}

\clearpage



\end{document}